\documentclass[aps,pra,twocolumn,superscriptaddress,showpacs,amsmath]{revtex4-1}

\usepackage{braket}
\usepackage[per=slash,
            emulate=units, 
            range-phrase = -,
            range-units = single,
            exponent-product = \cdot,
            inter-unit-product = \cdot,
            separate-uncertainty = true]{siunitx} 

\usepackage{hyperref}
\usepackage{graphicx}

\begin{document}

\title{Measurement and numerical calculation of Rubidium Rydberg Stark spectra}

\author{Jens Grimmel}
\email[]{jens.grimmel@uni-tuebingen.de} 
\affiliation{CQ Center for Collective Quantum Phenomena and their Applications, Physikalisches Institut, Eberhard-Karls-Universit\"at T\"ubingen, Auf der Morgenstelle 14, D-72076 T\"ubingen, Germany} 

\author{Markus Mack} 
\affiliation{CQ Center for Collective Quantum Phenomena and their Applications, Physikalisches Institut, Eberhard-Karls-Universit\"at T\"ubingen, Auf der Morgenstelle 14, D-72076 T\"ubingen, Germany} 

\author{Florian Karlewski} 
\affiliation{CQ Center for Collective Quantum Phenomena and their Applications, Physikalisches Institut, Eberhard-Karls-Universit\"at T\"ubingen, Auf der Morgenstelle 14, D-72076 T\"ubingen, Germany} 

\author{Florian Jessen} 
\affiliation{CQ Center for Collective Quantum Phenomena and their Applications, Physikalisches Institut, Eberhard-Karls-Universit\"at T\"ubingen, Auf der Morgenstelle 14, D-72076 T\"ubingen, Germany} 

\author{Malte Reinschmidt} 
\affiliation{CQ Center for Collective Quantum Phenomena and their Applications, Physikalisches Institut, Eberhard-Karls-Universit\"at T\"ubingen, Auf der Morgenstelle 14, D-72076 T\"ubingen, Germany} 

\author{N\'{o}ra S\'{a}ndor}
\affiliation{CQ Center for Collective Quantum Phenomena and their Applications, Physikalisches Institut, Eberhard-Karls-Universit\"at T\"ubingen, Auf der Morgenstelle 14, D-72076 T\"ubingen, Germany} 
\affiliation{Wigner Research Center for Physics, Hungarian Academy of Sciences, Konkoly-Thege Mikl\'os \'ut 29-33, H-1121 Budapest, Hungary}

\author{J\'{o}zsef Fort\'{a}gh}
\email[]{fortagh@uni-tuebingen.de}
\affiliation{CQ Center for Collective Quantum Phenomena and their Applications, Physikalisches Institut, Eberhard-Karls-Universit\"at T\"ubingen, Auf der Morgenstelle 14, D-72076 T\"ubingen, Germany} 

\date{\today}

\begin{abstract}
	We report on the measurement of Stark shifted energy levels of $^{87}$Rb Rydberg atoms in static electric fields by means of electromagnetically induced transparency (EIT). Electric field strengths of up to \SI{500}{\volt\per\centi\meter}, ranging beyond the classical ionisation threshold, were applied using electrodes inside a glass cell with rubidium vapour. Stark maps for principal quantum numbers $n=35$ and $n=70$ have been obtained with high signal-to-noise ratio for comparison with results from ab initio calculations following the method described in [M. L. Zimmerman et al., Phys. Rev. A 20, 2251 (1979)], which was originally only verified for states around $n=15$. We also calculate the dipole matrix elements between low-lying states and Stark shifted Rydberg states to give a theoretical estimate of the relative strength of the EIT signal. The present work significantly extends the experimental verification of this numerical method in the range of both high principal quantum numbers and high electric fields with an accuracy of up to \SI{2}{\mega\hertz}. 
\end{abstract}

\pacs{32.30.-r, 32.60.+i, 32.80.Ee} 

\maketitle

\section{\label{sec:intro}Introduction}

The response of atoms to static electric fields (DC Stark effect) results in line shifts, state mixing and, for sufficiently large fields, ionisation. The line shifts are conventionally summarised in Stark maps, displaying the energy levels as a function of the applied field. Stark maps of alkali atoms are routinely calculated by diagonalising the perturbed Hamiltonian \cite{Zimmerman.1979}, taking into account quantum defects and corresponding electronic wavefunctions \cite{Gallagher.1994}. A precise knowledge and control of Stark shifted Rydberg states is required for the application of Rydberg atoms as quantum probes \cite{Guerlin.2007}, for controlling the interactions between Rydberg atoms \cite{Daschner.2012}, the production of circular Rydberg atoms \cite{Raithel.2013}, the structure and dynamics of Rydberg gases \cite{Comparat.2010,Beterov.2009}, and possible applications in quantum information processing \cite{ParedesBarato.2014}. In the context of hybrid quantum systems based on atoms and solid state quantum circuits \cite{Petrosyan.2008,Tauschinsky.2010}, Stark shifts of Rydberg states \cite{Hattermann.2012,Abel.2011,Chan.2014} and their control \cite{Jones.2013} are of particular interest. 

Stark maps of Rydberg excited alkali atoms were studied in the 1970s using pulsed laser excitations and subsequent pulsed-field ionisation \cite{Littman.1976,Zimmerman.1979}. For low-lying Rydberg states of Sodium, Lithium and Caesium with principal quantum numbers $n<20$, Stark maps have been recorded up to and beyond the classical ionisation limit \cite{Littman.1976,Zimmerman.1979}. Stark maps of $^{85}$Rb for $n$ up to $55$ were studied in the 1980s for low electric fields using two-photon laser excitation and detecting ionisation from thermal collisions \cite{OSullivan.1986}. In this regime of low electric fields the Stark maps do not show level crossings but quadratic dependence on the applied field with slight deviation from this for highly excited states ($n=55$).

Many recent experiments on Stark shifts use electromagnetically induced transparency (EIT) \cite{Fahey.2011,Tauschinsky.2013}. This spectroscopic method provides a high resolution of the energy levels \cite{Tauschinsky.2013} and is suitable for the detection of states of high principal quantum numbers. However, the measurements so far only covered the range of low electric fields, in which just the first avoided crossings appear. Similarly, the detection of ions by micro-channel plates (MCP) \cite{Grabowski.2006} and the technique of measuring the ionisation currents from Stark-shifted Rydberg states \cite{Barredo.2013} have only been used at low electric fields. 

In this article, we report on the optical spectroscopy of Stark shifted Rydberg states with principal quantum numbers of $n=35$ and $n=70$ for electrostatic fields between \SIrange{0}{500}{\volt\per\centi\meter}, ranging beyond the classical ionisation limit. Our measurements go to higher principal quantum numbers and cover a three to four times larger range of electric fields relative to the classical ionisation threshold than any of the aforementioned works using EIT. The observed Stark maps are compared with results from numerical calculations following the lines of the numerical method by \cite{Zimmerman.1979}, including the recently improved accuracy of the quantum defects \cite{Mack.2011}. In the original work this method was optimised for Rydberg states around $n=15$ due to computational limitations as well as the accessible experimental data. The high accuracy of this method at low electric fields is always given by the accuracy of the unperturbed energy levels, which are used in the calculation. However, for strong electric fields and high principal quantum numbers, where a high accuracy of the calculated energy levels is desirable, this method becomes more susceptible to numerical errors \cite{Zimmerman.1979}. In this article we show by direct comparison, that it is still applicable even in these regions of study. Furthermore, we calculate dipole matrix elements between $5\text{P}_{3/2}$ and the observed Stark shifted states, which are then used to give an estimate for the relative strength of the measured signals.

\section{\label{sec:exp}Measurement of Stark maps} 

\begin{figure}
	\includegraphics{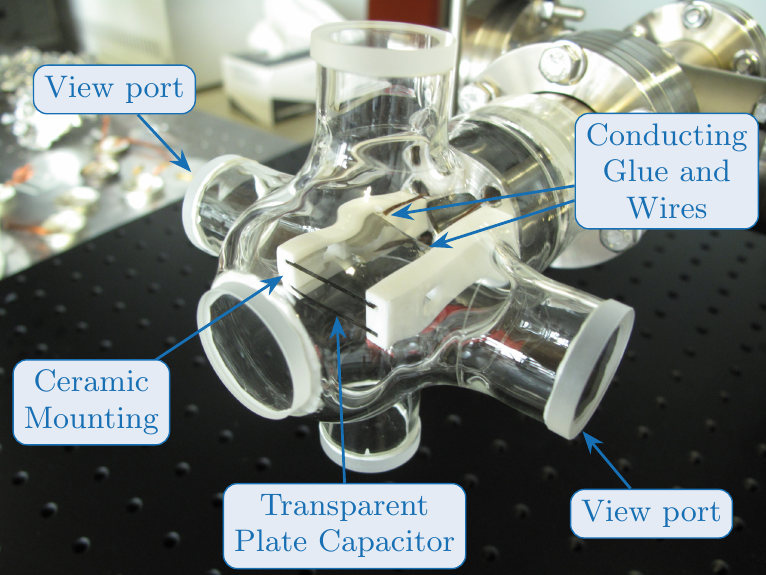}
	\caption{\label{fig:photo_cell_nice} (Color online) Vapour cell with electrodes. The capacitor plates are formed by two glass plates with \SI{5}{\nano\meter} of Nickel facing each other at a separation of \SI{5}{\milli\meter}. Two wires are glued to each capacitor plate to contact them. For the measurements the laser beams counterpropagate through the view ports indicated in the image.}
\end{figure}

For the measurement of Stark shifts of $^{87}$Rb Rydberg atoms we use a vapour cell with a pair of plate electrodes for applying homogeneous electric fields. The electrodes are inside the cell [Fig. \ref{fig:photo_cell_nice}], trying to avoid the effect observed in previous works where static electric fields applied to a vapour cell with outside electrodes are compensated by ionised Rubidium and electrons \cite{Mohapatra.2007}. The Rubidium vapour in the cell is at room temperature with an estimated pressure of \SI{\approx1e-7}{\milli\bar}. The electrodes in the center of the glass cell [Fig. \ref{fig:photo_cell_nice}] are formed by two square glass plates coated with \SI{5}{\nano\meter} of Nickel. This gives the plates \SI{\approx60}{\percent} transparency, which allows for optical access on the axis perpendicular to them. However, this optical access was not used in the experiments presented in this work. The plates are mounted with \SI{5}{\milli\meter} separation on insulating ceramics (Macor) and connected to a voltage source through metallic wires. Laser beams for the optical spectroscopy are introduced through view ports and pass through the cell between the plates. 

\begin{figure} 
	\includegraphics{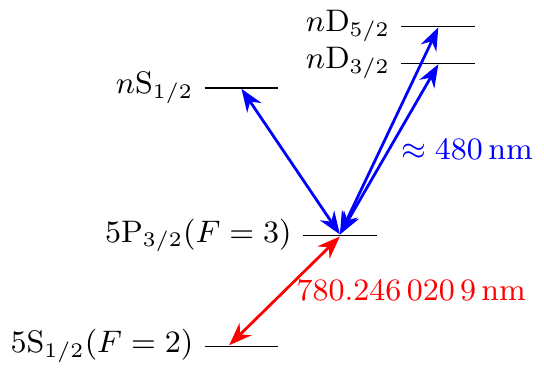}
	\caption{\label{fig:anregung} (Color online) Ladder scheme for electromagnetically induced transparency (EIT). For our measurements the probe laser is held on resonance with the $5\text{S}_{1/2}(F{=}2)\rightarrow5\text{P}_{3/2}(F{=}3)$ transition of $^{87}$Rb and the coupling laser is scanned around the upper transition. Dipole selection rules only allow for the measurement of $n\text{S}$ and $n\text{D}$ states in this scheme.}
\end{figure}

We measure the transition frequency from the Rubidium ground state to Rydberg states by electromagnetically induced transparency (EIT). The EIT three level ladder scheme \cite{Mohapatra.2007} consists of the ground state $5\text{S}_{1/2}(F{=}2)$, the intermediate state $5\text{P}_{3/2}(F{=}3)$ and a Rydberg state $n\text{S}$ or $n\text{D}$ [Fig. \ref{fig:anregung}]. However, it is important to keep in mind that $l$ is not a good quantum number any more in the presence of an external electric field and that the Stark shifted states can be considered as a mix of all possible unperturbed $l$ states. For the spectroscopy we use a \textit{probe} laser that is locked to the $5\text{S}_{1/2}(F{=}2) \rightarrow 5\text{P}_{3/2}(F{=}3)$ transition (\SI{780}{\nano\meter}) and a \textit{coupling} laser with a variable frequency close to the transition between $5\text{P}_{3/2}(F{=}3)$ and a Rydberg state (\SI{480}{\nano\meter}). When the coupling laser is resonant with this transition we detect a maximum in the transmission of the probe laser through the vapour. 

\begin{figure}
	\includegraphics{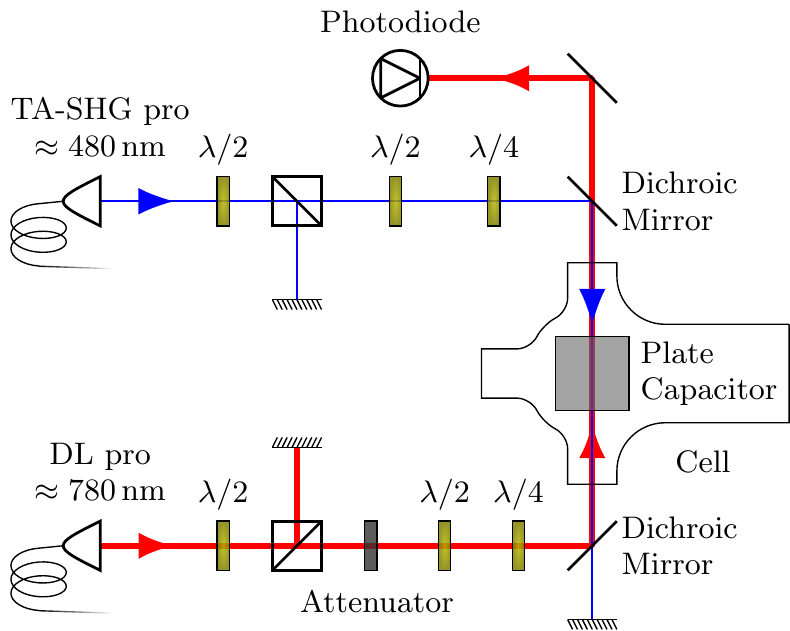}
	\caption{\label{fig:aufbau_schema} (Color online) Optical setup (schematic) for the measurement of Stark maps using EIT. Both lasers are guided to the experiment through optical fibres and overlapped within the volume of the plate capacitor using two dichroic mirrors. Intensity and polarisation of both lasers are adjusted independently. The transmission of the probe beam through the vapour cell is measured with a photodiode.} 
\end{figure}

We use a grating-stabilised diode laser (Toptica, DL pro) of \SI{\approx100}{\kilo\hertz} linewidth as the probe laser and a frequency-doubled, grating-stabilised diode laser of \SI{\approx200}{\kilo\hertz} linewidth (Toptica, TA-SHG pro) as the coupling laser. Both lasers are phase-locked to a frequency comb (Menlo Systems, FC 1500). For conveniently selecting the right modes of the frequency comb for both lasers we use a calibrated wavelength meter (HighFinesse, WS Ultimate 2) \cite{Mack.2011}. The power of the probe laser is \SI{1}{\micro\watt} and the power of the coupling laser is \SI{25}{\milli\watt} with $1/e^2$ diameters of \SI{450}{\micro\meter} and \SI{150}{\micro\meter} in the cell, respectively. The small diameter of the coupling beam results in a high intensity and therefore high Rabi frequency on the corresponding transition while the bigger diameter of the probe beam is chosen in order to ensure maximal overlap of the two beams within the cell. The laser powers and polarisations were adjusted to maximise the EIT signal in zero field [Fig. \ref{fig:aufbau_schema}]. A measurement of the polarisations after this adjustment revealed that this resulted in circular polarisations for both lasers. The frequency of the probe laser is kept on resonance with the $5\text{S}_{1/2}(F{=}2)\rightarrow5\text{P}_{3/2}(F{=}3)$ transition [Fig. \ref{fig:anregung}] and its transmission through the cell is measured using an avalanche photodiode. 

In order to reach a high signal-to-noise ratio in the spectroscopy, we use the following lock-in measurement technique. We modulate the power of the coupling laser using an acousto-optic modulator (AOM) and demodulate the signal from the photodiode on the carrier frequency. In addition we modulate the frequency of the intensity modulation and demodulate the photodiode signal on two of the sidebands that arise from this. We then average the demodulated signal from the carrier and both sidebands. The additional frequency modulation decreases the signal strength on the carrier frequency and creates an even lower signal strength on the sidebands, but averaging the signal from three frequencies lowers the noise level at the same time. In total, the resulting signal-to-noise ratio is improved by a factor of \num{\approx2} as compared to a simple lock-in measurement without frequency modulation. For every set frequency of the coupling laser we ramp up the voltage on the plate capacitor using an auxiliary output of the lock-in amplifier (Zurich Instruments, HF2LI). This is sufficient for electric fields up to \SI{\approx20}{\volt\per\centi\meter}. For even higher fields up to \SI{\approx500}{\volt\per\centi\meter} we use an additional voltage amplifier. 

\begin{figure*}
	\includegraphics{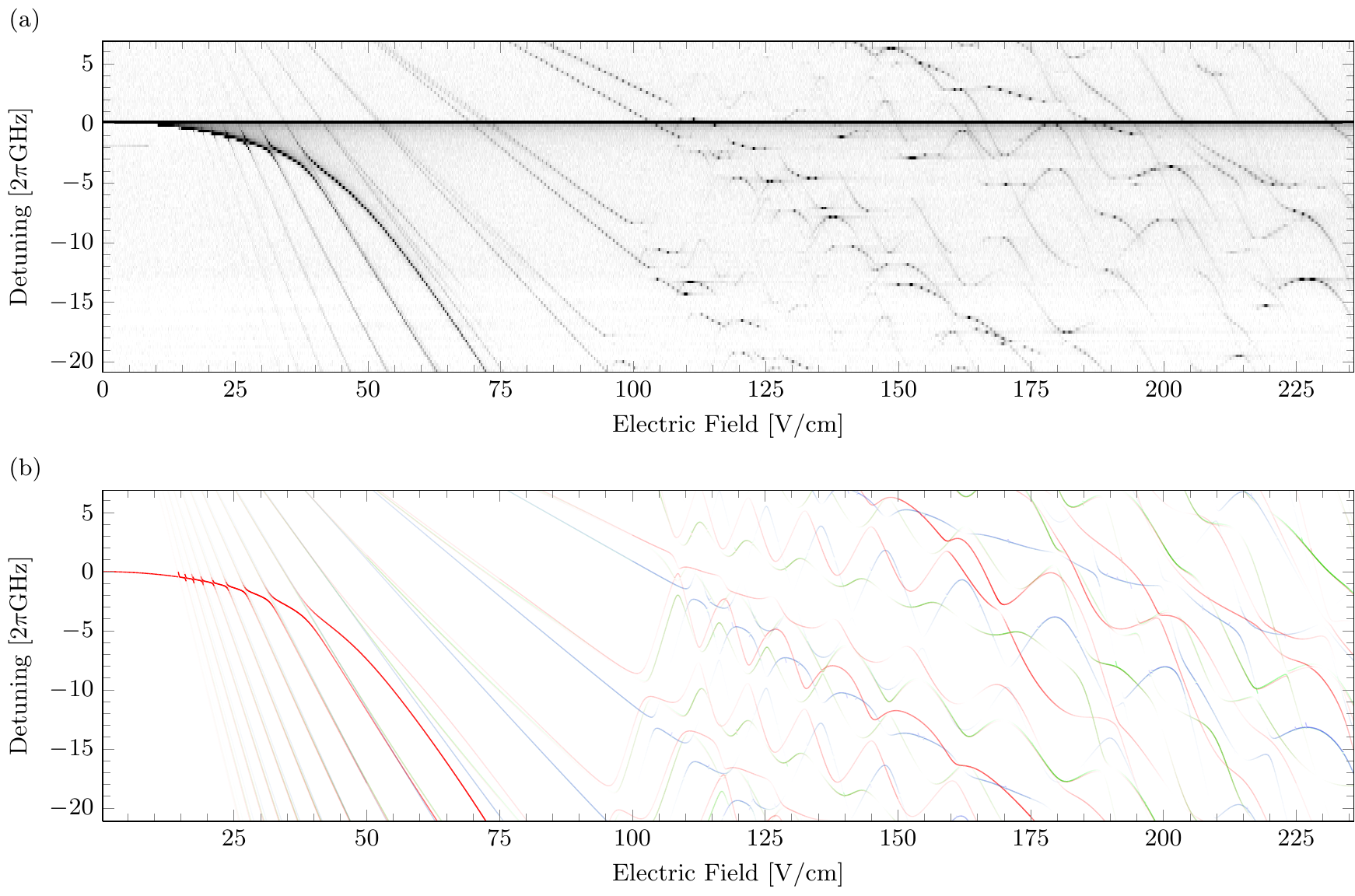}
	\caption{\label{fig:35S-1-2-th-exp-a} (Color online) (a) Stark map for $35\text{S}_{1/2}$ for an electric field range of \SIrange{0}{235}{\volt\per\centi\meter}. The gray scale is determined from the measured EIT signal, which represents the demodulated transmission signal from the photodiode. Due to the frequency resolution of \SI{125}{\mega\hertz} some of the lines with a small slope appear dotted here. (b) The red, green and blue lines show the numerically calculated Stark map for $\left|m_j\right|=1/2$, $\left|m_j\right|=3/2$ and $\left|m_j\right|=5/2$ respectively. The opacity gradients indicate the calculated transition strength $D$ to the corresponding Stark shifted Rydberg states. In this region we find a very good agreement between experimental and numerically calculated results in the Stark shifts as well as the transition strengths.}
\end{figure*}

\begin{figure*}
	\includegraphics{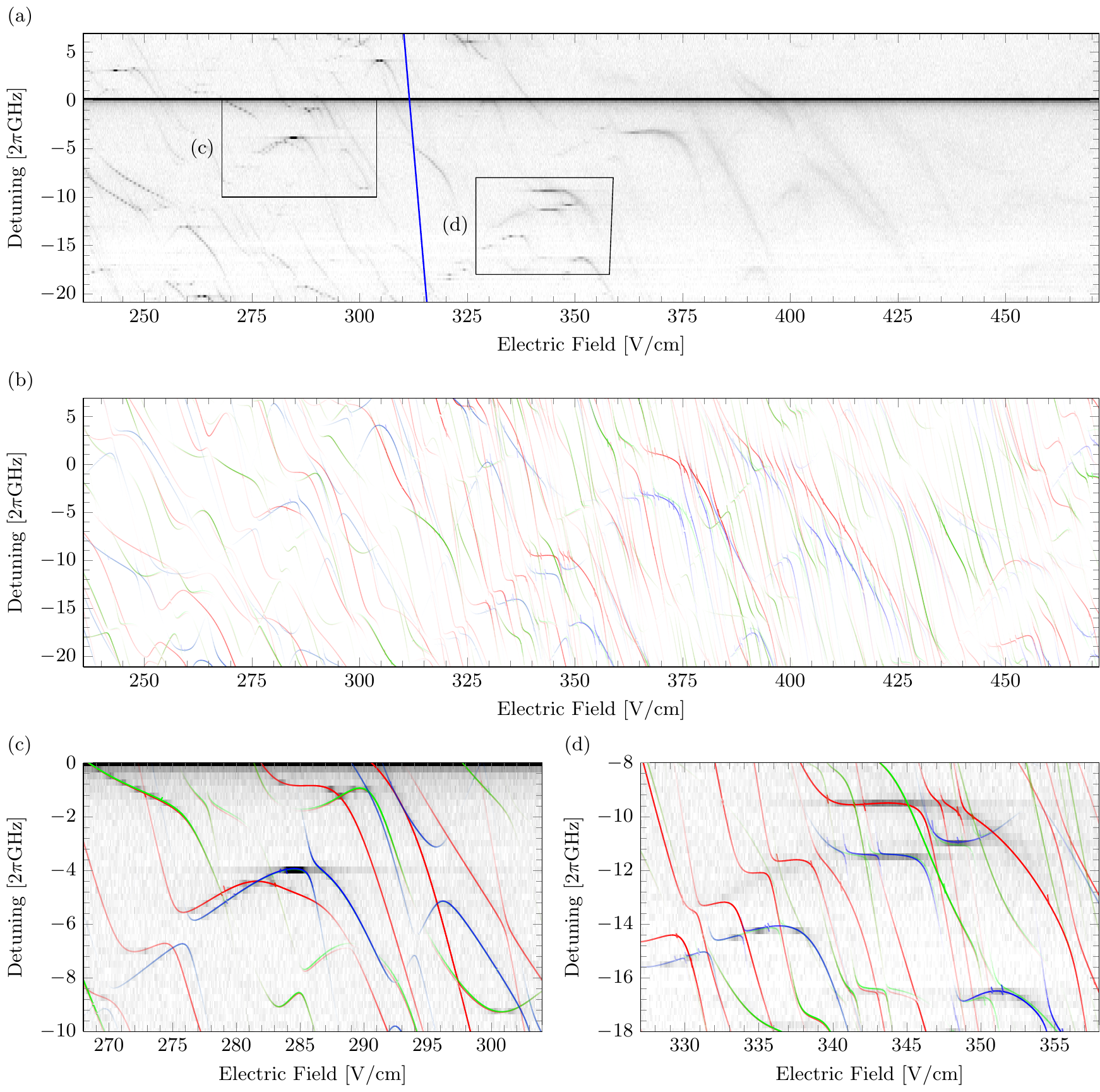}
	\caption{\label{fig:35S-1-2-th-exp-b} (Color online) (a) Stark map for $35\text{S}_{1/2}$ for an electric field range of \SIrange{235}{470}{\volt\per\centi\meter}. The gray scale represents the measured EIT signal. Due to the frequency resolution of \SI{125}{\mega\hertz} some of the lines with a small slope appear dotted here. The blue solid line indicates the classical ionisation limit. Beyond this limit fewer states are present in the measurement. The states also appear broader and weaker as the external electric field lowers and extends the opening of the potential barrier it forms with the atomic potential. (b) Results from our numerical calculations for the same region. The red lines represent $\left|m_j\right|=1/2$, the green lines $\left|m_j\right|=3/2$ and the blue lines $\left|m_j\right|=5/2$ with the opacity gradients indicating the calculated transition strengths $D$ to each state. (c) and (d) show details of the measurement with the numerically calculated Stark map on top. The opacity of the colors has been increased with a linear scaling in comparison to (b) for better visibility of the calculated lines on top of the experimental data. The agreement in the energy levels remains even beyond the classical ionisation threshold, but we find some discrepancies in the transition strengths as the electric field strength increases.}
\end{figure*}

\begin{figure}
	\includegraphics{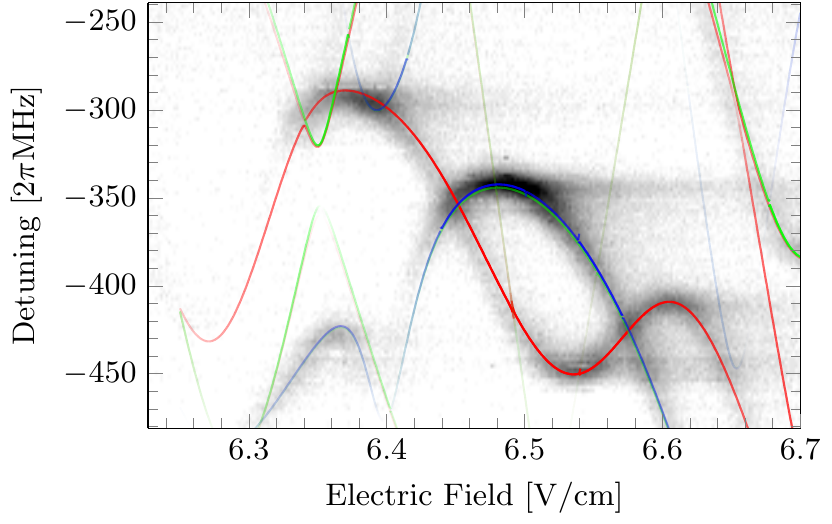}
	\caption{\label{fig:70S-1-2-th-exp} (Color online) Detailed part of a Stark map for $70\text{S}_{1/2}$ for an electric field range of \SIrange{6.23}{6.70}{\volt\per\centi\meter}. The gray scale represents the measured EIT signal. The red, green and blue lines plotted on top of the measured signal show the numerically calculated Stark map for $\left|m_j\right|=1/2$, $\left|m_j\right|=3/2$ and $\left|m_j\right|=5/2$ respectively with the opacity gradient scaled to the calculated transition strength $D$. The frequency resolution of this measurement is \SI{2}{\mega\hertz} and, again, we find a very good agreement between the experimental and numerically calculated results.} 
\end{figure}

Measured data near the unperturbed $35\text{S}_{1/2}$ state is shown in Figs. \ref{fig:35S-1-2-th-exp-a} and \ref{fig:35S-1-2-th-exp-b}. Near the unperturbed $70\text{S}_{1/2}$ state we conducted a preliminary measurement, which is not shown in this work, and then selected a smaller region for a more detailed measurement in order to test the frequency precision of the numerical calculations [Fig. \ref{fig:70S-1-2-th-exp}]. Each pixel in the gray scale images represents an average of the demodulated signal over \SI{200}{\milli\second}. Between every two pixels we add a waiting time of \SI{50}{\milli\second} to allow for the low-pass filter of the lock-in amplifier to settle. The gray scale was adjusted with a cutoff for better visibility of weaker signals. Our data shows states ranging up to and even beyond the classical ionisation threshold that have not been measured by means of EIT before. The classical ionisation threshold $E_\text{ion}$, i.e. the saddle point which is formed by a Coulomb potential with an external electric field $F$, is given by 
\begin{equation}
	E_\text{ion} = - 2 \sqrt{F}
\end{equation}
in atomic units \citep{Gallagher.1994}. This results in an electric field strength for the ionisation threshold of \SI{\approx312}{\volt\per\centi\meter} for $35\text{S}_{1/2}$ and \SI{\approx16}{\volt\per\centi\meter} for $70\text{S}_{1/2}$. For a quantitative analysis we give a brief review of the numerical calculation of Stark maps in section \ref{sec:th}. 

We observe two background effects [Fig. \ref{fig:35S-1-2-th-exp-a}, \ref{fig:70S-1-2-th-exp}] which are caused by the region of the cell that is not covered by the plate capacitor [Fig. \ref{fig:photo_cell_nice}]. The first is the line of the unperturbed state, which remains visible for all applied voltages because we probe those outer regions of the cell as well. The second is a smearing of the lines to the right at low fields as visible in Fig. \ref{fig:35S-1-2-th-exp-a} at the avoided crossings up to \SI{\approx50}{\volt\per\centi\meter} and in Fig. \ref{fig:70S-1-2-th-exp}. For a certain electric field strength inside the plate capacitor one always finds lower electric field strengths in the inhomogeneous outside region, causing the asymmetry of the smearing to the right. 

Inside the capacitor undesirable electric fields could also arise from the dipole which is formed between adsorbed Rubidium and the Nickel surface of the capacitor plates. In previous experiments the repeated deposition of cold atom clouds of Rubidium onto a Copper surface led to electric fields close to the surface which saturated as the number of deposited clouds increased \cite{Hattermann.2012}. Since the work functions of Nickel and Copper are similar and the cell is filled with Rubidium vapour at all times we may find a similar effect for the capacitor plates. This effect may play a role for Rydberg states higher than $n=70$ at low electric fields, but is negligible for the measured data presented in this article, where the distance between the laser beams and the capacitor plates is \SI{\approx2}{\milli\meter}. Other stray fields could originate from outside of the cell, but should be compensated by electrons and ionised Rubidium \cite{Mohapatra.2007}. 

\section{\label{sec:th}Calculation of Stark maps and dipole matrix elements}

The Hamiltonian for the valence electron in an alkali atom with an external electric field in z-direction can be written as 
\begin{equation}
	\hat{H} = \hat{H}_0 + E\hat{z}
\end{equation} 
in atomic units with $\hat{H}_0$ the Hamiltonian for the valence electron in absence of any perturbation, $E$ the electric field strength and $\hat{z}$ the position operator in z-direction. For the following calculations we include the fine structure splitting in $\hat{H}_0$ and neglect the hyperfine structure splitting as it is smaller than our frequency resolution in the experiment \cite{Zimmerman.1979}. The Stark shifts of the $5\text{S}_{1/2}(F{=}2)$ and $5\text{P}_{3/2}(F{=}3)$ states from the EIT scheme are negligible as well, as they amount to less than \SI{100}{\kilo\hertz} for the region of the field strength up to \SI{500}{\volt\per\centi\meter}. We then create a matrix representation of $\hat{H}$ in a subset of the basis given by $\hat{H}_0$. This way $\hat{H}_0$ is represented by a diagonal matrix with the energy levels from \cite{Mack.2011} on its diagonal. 

The matrix representation of $\hat{z}$ is symmetric with only off-diagonal entries and in spherical coordinates we obtain from \cite{Zimmerman.1979} 
	\begin{align} 
		\label{equ:z}
		& \Braket{n,l,j,m_j|\hat{\text{z}}|n',l',j',m'_j} \nonumber\\ 
			& \quad = \delta_{m_j,m'_j}\delta_{l,l'\pm1}\Braket{n,l,j|r|n',l',j'} \nonumber\\ 
			& \qquad \times \sum\limits_{m_l=m_j\pm\frac{1}{2}} \left\lbrace\Braket{l,\frac{1}{2},m_l,m_j-m_l|j,m_j}\right. \nonumber\\ 
		& \qquad \times \Braket{l',\frac{1}{2},m_l,m_j-m_l|j',m_j} \nonumber\\ 
		& \qquad \times \left. \Braket{l,m_l|\cos\theta|l',m_l} \vphantom{\frac{1}{2}} \right\rbrace  
	\end{align} 
with the radial overlap integral $\Braket{n,l,j|r|n',l',j'}$ in the first line of the right hand side. From the Kronecker delta $\delta_{m_j,m'_j}$ we can see that the matrix representations of $H$ for different values of $m_j$ can be calculated separately. It is also evident that the Stark shifts for $\pm|m_j|$ are always equal. Therefore we only calculate matrix representations for positive values of $m_j$ in this step. This reduces the computing time for diagonalising the matrices, but we need to consider states with both signs for the calculation of transition strengths later on again. 

For the calculation of the overlap integrals we tried two different methods to calculate the radial wavefunctions for all states in the chosen subset of the basis and also for $5\text{S}_{1/2}(F{=}2)$ and $5\text{P}_{3/2}(F{=}3)$, which will be used later on in the calculation of the measure for the transition strengths. One method was to further follow \cite{Zimmerman.1979} using their atomic potential for $^{87}$Rb and Numerov's method to solve the radial Schr\"{o}dinger equation. We also tried using the parametric model potential from \cite{Marinescu.1994} with another solving algorithm \cite{ode45}, which allowed us to obtain parts of the wavefunctions that are located further inside the ionic core. While the two methods lead to slightly different amplitudes of the wavefunctions due to normalisation, the differences in the Stark shifts and transition strengths calculated from the two methods for the observed states lie below our experimental accuracy. All calculations presented in this work utilised the latter method.

We can then calculate the matrix representations of the perturbed Hamiltonians $H$ using equation \eqref{equ:z} and diagonalise them efficiently in parallel for different values of the electric field strength to obtain the energy eigenvalues. In this step we also calculate the eigenvector $\beta$ corresponding to every eigenvalue. With this we can further follow the method from \cite{Zimmerman.1979} to represent every Stark shifted Rydberg state $\xi$ as a linear combination of unperturbed states, given by the eigenvectors, and calculate the dipole matrix elements 
\begin{align}
	&\Braket{\xi|\vec{r}|5\text{P}_{3/2}, F{=}3, m_F} \nonumber\\ 
		&\quad = \sum\limits_{n,l,j,m_j} \beta_{n,l,j,|m_j|} \Braket{n,l,j,m_j|\vec{r}|5\text{P}_{3/2}, F{=}3, m_F} 
\end{align} 
with $\beta_{n,l,j,|m_j|}$ the entry of $\beta$ that corresponds to the states $\Ket{n,l,j,\pm |m_j|}$. Here we sum over all states in the selected subset of the basis given by the unperturbed states and both signs for each value of $m_j$. For the calculation of the dipole matrix elements between the unperturbed Rydberg states and $\Ket{5\text{P}_{3/2}, F{=}3, m_F}$ we separate the radial and the angular parts of the wavefunctions. We use the radial overlap integrals from equation \ref{equ:z} and calculate the three components of the angular overlap integrals following \cite{Bethe.1957}. 

We take into account the effect of the probe laser, which couples the different $m_F$ substates of $\Ket{5\text{S}_{1/2}, F{=}2}$ and $\Ket{5\text{P}_{3/2}, F{=}3}$ by calculating the weighting factors 
\begin{align}
	\eta_{m_F} = \sum\limits_{m'_F} \big|\Braket{5\text{P}_{3/2}, F{=}3, m_F | \vec{\varepsilon}_{\text{p}}\vec{r} | 5\text{S}_{1/2}, F'{=}2, m'_F}\big|^2
\end{align}
with $\vec{\varepsilon}_{\text{p}}$ the polarisation of the probe laser. In this step we assume that all $m_F$ substates of $\Ket{5\text{S}_{1/2}, F{=}2}$ are evenly occupied. The weighting factors are then used to calculate the measure for the transition strength 
\begin{equation}
	\label{equ:transitionstrength}
	D = \sum\limits_{m_F} \eta_{m_F} \big|\Braket{\xi|\vec{\varepsilon}_{\text{c}}\vec{r}|5\text{P}_{3/2}, F{=}3, m_F}\big|^2
\end{equation}
with $\vec{\varepsilon}_{\text{c}}$ the polarisation of the coupling laser. In the experiment the two lasers are counterpropagating and perpendicular to the external electric field with circular polarisations. Therefore we use $\vec{\varepsilon}_p=\vec{\varepsilon}_c=(0,i,1)$ here. The measure for the transition strength $D$ is used to determine the opacity gradients for the lines in Fig. \ref{fig:35S-1-2-th-exp-a}, \ref{fig:35S-1-2-th-exp-b} and \ref{fig:70S-1-2-th-exp}. 

The Stark map for $35\text{S}_{1/2}$, which is shown in Figs. \ref{fig:35S-1-2-th-exp-a} and \ref{fig:35S-1-2-th-exp-b}, was calculated from a subset of the basis of approximately \num{1600} states and for \num{2000} values of the electric field strength using MATLAB. For $70\text{S}_{1/2}$, shown in Fig. \ref{fig:70S-1-2-th-exp}, the calculations run similarly with \num{4000} states and \num{500} values of the electric field strength. The number of states that was used for these calculations was adjusted so that a further increase only yields changes which lie below the accuracy of the experimental data. Calculated Stark maps for $\left|m_j\right|=1/2$, $\left|m_j\right|=3/2$ and $\left|m_j\right|=5/2$ are included in the figures. The measurements were performed in frequency regions around S states, but since $l$ is not a good quantum number any more in the presence of an electric field \cite{Zimmerman.1979}, we find that other states with $\left|m_j\right|=3/2$ and $\left|m_j\right|=5/2$ are shifted far enough by the Stark effect to appear within the measured frequency range.

\section{\label{sec:results}Comparison of Measurements and Calculations}

Figs. \ref{fig:35S-1-2-th-exp-a} and \ref{fig:35S-1-2-th-exp-b} show comparisons of the measured and numerically calculated Stark maps for $35\text{S}_{1/2}$. The frequency axis shows the detuning of the coupling laser relative to the absolute value of the transition frequency from \cite{Mack.2011}. A linear scaling with an offset has been applied to the electric field axis of the measured data. Using only this scaling we achieve a match between calculated and measured energy levels for Stark states in the whole range of our measurement. 

The lines in the calculated Stark maps are drawn with an opacity gradient, resulting in a color range between white and the respective color associated with the different values for $|m_j|$. The opacity gradient is scaled to $D$ from equation \eqref{equ:transitionstrength} with an upper cutoff at \SI{70}{\percent} of its maximum for better visibility. We find a good agreement between this calculated measure for the transition strength and the experimental data in the range of electric fields shown in Fig. \ref{fig:35S-1-2-th-exp-a}. The quality of the agreement deteriorates slightly approaching the classical ionisation threshold, but there is still good agreement beyond this point, as can be seen in Fig. \ref{fig:35S-1-2-th-exp-b}. The very weak signals we still find in the experimental data at electric fields around \SI{400}{\volt\per\centi\meter} all correspond to calculated states with a strong calculated transition strength. Considering this, we think that the method described here can be applied for these high electric field strengths, where only few states are still visible in the experimental data, as a way to select potentially interesting areas before a measurement. 

One important cause for discrepancies between the experimental and numerically calculated data we present here is that no ionisation effects were taken into account for the calculations. However, it is interesting to note that even though the calculated measure for the transition strength shows some differences at high electric fields, we still find a remarkable agreement of calculated and measured energy levels in this region of study. These differences could be related to the calculation of the radial wavefunctions. We calculate the radial wavefunctions for $\Ket{5\text{S}_{1/2}}$ and $\Ket{5\text{P}_{3/2}}$ using the same method as for Rydberg states. Evidently this works quite well here, but more accurate radial wavefunctions, especially for $\Ket{5\text{P}_{3/2}}$ but also for Rydberg states with low $l$, could further improve the quality of the calculated transition strengths. This is further emphasised as the biggest discrepancies of the transition strengths at high electric fields can be found for states $|m_j|=1/2$ while states with $|m_j|=5/2$ show hardly any discrepancies on the whole range of electric fields. On the other hand, these discrepancies could also be explained by a stronger coupling of Rydberg states with low $l$ to the continuum. Some disagreements on the whole range of our measurements can also be caused by the scaling of the EIT signal with the dipole matrix elements, which is only approximated by the squared scaling in $D$. Furthermore, lines with small slopes and especially local extrema may appear over-pronounced in the experimental data. Since these parts of the lines are broad in the electric field domain and the inhomogeneous electric field outside of the plate capacitor still lies within the volume of the cell, this results in an effective contribution of more atoms to the signal. 

We find a match similar to the one presented for $35\text{S}_{1/2}$ in the energy levels and transition strengths for $70\text{S}_{1/2}$, which is shown in detail in Fig. \ref{fig:70S-1-2-th-exp}. Discrepancies lie within the frequency resolution of \SI{2}{\mega\hertz} for Fig. \ref{fig:70S-1-2-th-exp}, which is the highest accuracy we present here. The opacity gradients were assigned the same way as for $35\text{S}_{1/2}$. The calculated transition strengths match the experimental data equally well as for $35\text{S}_{1/2}$. The experimental data presented in Figs. \ref{fig:35S-1-2-th-exp-a}, \ref{fig:35S-1-2-th-exp-b} and \ref{fig:70S-1-2-th-exp} is provided as supplementary data to this article \cite{supp_mat}. All results presented in this work were measured and calculated in the energy range near $n\text{S}$ states. However, they intrinsically include contributions all possible $l$ states due to the mixing of states caused by the external electric field. 

For a possible extension of our measurements to a wider range of principal quantum numbers $n$ than the region between $n=35$ and $n=70$, the most important limiting factor is that the coupling of Rydberg states to the intermediate $5\text{P}_{3/2}(F{=}3)$ grows weaker as $n$ increases, which leads to a lower signal-to-noise ratio. Another factor influencing the signal-to-noise ratio is the density of the vapour, which on the other hand gives rise to collective effects which in turn cause undesired changes of the EIT signal \cite{Ates.2011}.

\section{\label{sec:conclusion}Conclusion}

In summary, we have demonstrated agreement between measured and calculated Stark maps of $^{87}$Rb up to an accuracy of \SI{2}{\mega\hertz}. This agreement holds for the range from zero field to beyond the classical ionisation threshold and for principal quantum numbers $n=35$ and $n=70$. Such data and calculations may aid the accurate mapping of electric fields at surfaces \cite{Hattermann.2012}. Furthermore, we presented numerical calculations to estimate the transition strength from low-lying states to Stark shifted Rydberg states. The results from these calculations show a very good agreement with our experimental data on a high range of electric fields and even beyond the classical ionisation threshold. 

Altogether, the methods presented in this paper can be used to find experimentally accessible Stark shifted Rydberg states with an appropriate sensitivity to external electric fields for a wide range of applications. For example, the dependence of the transition strength to different Stark shifted Rydberg levels could be used to determine not only the strength of an external electric field in an experimental system but also the electric field axis.

\appendix* 

\begin{acknowledgments}
	This work was financially supported by the FET-Open Xtrack Project HAIRS and the Carl Zeiss Stiftung. N\'{o}ra S\'{a}ndor acknowledges financial support from the framework of T\'{A}MOP-4.2.4.A/2-11/1-2012-0001 'National Excellence Program'. We thank Nils Schopohl and Ali Sanayei for helpful discussions. 
\end{acknowledgments}

\end{document}